\documentclass[aps,groupaddress,showpacs,preprint]{revtex4}
\usepackage[dvips]{graphicx}

\begin{document}

\title{Anderson localization of electron states in graphene in
different types of disorder}

\author{Shi-Jie Xiong}
\email{sjxiong@nju.edu.cn}
\address{National Laboratory of Solid State Microstructures and
Department of Physics, Nanjing University, Nanjing 210093, China}
\author{Ye Xiong}
\affiliation{College of Physical Science and Technology, Nanjing
Normal University, Nanjing 210097, China}

\begin{abstract}

Anderson localization of electron states on graphene lattice with
diagonal and off-diagonal (OD) disorder in the absence of magnetic
field is investigated by using the standard finite-size scaling
analysis. In the presence of diagonal disorder all states are
localized as predicted by the scaling theory for two-dimensional
systems. In the case of OD disorder, the states at the Dirac point
($E=0$) are shown to be delocalized due to the specific chiral
symmetry, although other states ($E \neq 0$) are still localized. In
OD disorder the conductance at $E=0$ in an $M\times L$ rectangular
system at the thermodynamical limit is calculated with the
transfer-matrix technique for various values of ratio $M/L$ and
different types of distribution functions of the OD elements
$t_{nn'}$. It is found that if all the $t_{nn'}$'s are positive the
conductance is independent of $L/M$ as restricted by 2 delocalized
channels at $E=0$. If the distribution function includes the sign
randomness of elements $t_{nn'}$, the conductivity, rather than the
conductance, becomes $L/M$ independent. The calculated value of the
conductivity is around $\frac{4e^2}{h}$, in consistence with the
experiments.

\end{abstract}


\maketitle


\section{Introduction}

As a potential candidate material for carbon-based electronics,
graphene, a two-dimensional (2D) honeycomb lattice of carbon atoms,
has attracted much attention \cite{1,2,3,4,5}. It is found that both
the type (electrons or holes) and the number of carriers can be
tuned by changing the gate voltage \cite{6,7}. Such tunability
originates from the specific dispersion relation of the massless
Dirac fermions in graphene \cite{8a,8}. Its transport properties are
very special. There exists a universal maximal resistivity for all
samples with the Fermi level near the Dirac point, independent of
their shapes and mobility. The maximal resistivity \cite{3} implies
the existence of channels of extended states at the thermodynamical
limit, and the shape independence of the resistivity suggests that
both the number and spatial configuration of these extended channels
depend on the shape. This behavior is certainly exotic as from the
scaling theory all the states are localized in 2D systems \cite{9}.
The non-zero minimal conductivity has been investigated by several
theoretical studies using different methods \cite{a5,a6,a7,a8,a9}.
The minimum conductivity was correctly demonstrated, but as to the
reason for it and as to its exact value being $4e^2/h$ or $4e^2/\pi
h$ there is no consensus at present. Morozov {\it et al.} have found
that in graphene the weak localization is strongly suppressed and
they attributed this to mesoscopic corrugations of graphene sheets
which can cause a dephasing effect similar to that of a random
magnetic field \cite{morozov}. Theoretically, the influence of
disorder on 2D electron gases on the honeycomb lattice was studied
with a self-consistent Born approximation \cite{born1,born2}.
Pereira {\it et al.} have investigated the localized states near
vacancies in graphene \cite{pereira}. Khveshchenko has carried out
an analysis of quantum interference effects in disordered graphene
and specified the conditions for the quantum correction to the
conductivity to be positive, negative, and zero \cite{khv}. By
evaluating the dependence of the magnetoresistance of graphene on
the relaxation rates from different sources McCann {\it et al.}
showed that the warping disorder and the disorder due to atomically
sharp scatterers have different effects on the magnetoresistance
\cite{mccann}. Morpurgo and Guinea investigated the effects of
static disorder caused by curvature or topological defects and found
that when the intervalley scattering time is long enough an
effective time-reversal symmetry breaking can be induced
\cite{morpurgo}. Aleiner and Efetov showed two different types of
logarithmic contributions to conductivity by describing the disorder
with a renormalization-group equation \cite{aleiner}. From a
combination of mean-field and bosonization methods Altland pointed
out that there are several mechanisms conspiring to protect the
conductivity although disordered graphene is subject to the common
mechanisms of Anderson localization \cite{altland}. From the linear
response theory Ziegler indicated that the transport properties in
graphene are dominated by diffusion \cite {ziegler}. Based on
numerical study with finite-size Kubo formula for doped graphene,
Nomura and MacDonald found distinct differences between the cases of
short-range and Coulomb randomly distributed scatterers and
speculated that the transport properties in graphene are related to
the Boltzmann theory for Coulomb scatterers \cite{nomura}. Sheng
{\it et al}. carried out transfer matrix studies of graphene in a
magnetic field to investigate the disorder effect and phase diagram
\cite{sheng}. Louis {\it et al}. calculated conductances for samples
with edge disorder and obtained size independent conductivity as
$\sim 4e^2/\pi h$ at $E=0$ \cite{louis}. Moreover, by adopting
disorder model without intervalley scattering, Ostrovsky {\it et
al}. constructed an effective field theory and showed that the
system is at a quantum critical point with a universal value of the
conductivity of the order of $e^2/h$ \cite{Ostrovsky}. With a
similar model Titov demonstrated the impurity-assisted tunneling by
using the transfer matrix approach \cite{titov}. Bardarson {\it et
al}. numerically showed that the conductivity at the Dirac point
increases logarithmically with sample size in the absence of
intervalley scattering \cite{Bardarson}. From numerical calculation
Nomura {\it et al}. demonstrated that all states are delocalized
even in the strong disorder regime if the intervalley scattering is
excluded \cite{nomura2}. San-Jose {\it et al}. investigated the full
transport statistics of graphene at low dopings without intervalley
scattering and found identical scaling of all current cumulants
\cite{San-Jose}.

In this paper we investigate the Anderson localization of fermions
at the Dirac point in graphene with different types of disorder by
using the standard finite-scaling analysis together with the
transfer-matrix technique. The Dirac point corresponds to the band
center ($E=0$) of a tight-binding Hamiltonian on the honeycomb
lattice. In the real-space tight-binding representation the disorder
can be introduced with randomly distributed site energies, as called
diagonal disorder, or with random hopping integrals named as
off-diagonal (OD) disorder. It is known that the localization
behavior at the band center of a bipartite lattice with OD disorder
is anomalous due to the chiral symmetry \cite{a11,a12,a13}. For
instance, the decay of wavefunctions at the band center on 2D square
lattice with OD disorder is much weaker than exponential. The
honeycomb lattice is also bipartite, and there are other unusual
features at the Dirac point such as the massless Dirac fermion
dispersion relation. So it is interesting to investigate the
difference in the localization behavior at the Dirac point between
diagonal and off-diagonal disorders. We derive the formula of
transfer matrix for honeycomb lattice from which the numerical
studies can be carried out to calculate interesting properties such
as the rescaled localization length, the size and shape dependencies
of the conductance. The finite-size scaling analysis reveals that
although all the states are localized in the system with diagonal
disorder, in the case of OD disorder, the states at the Dirac point
are delocalized. From the calculated shape dependence of the
conductance at $E=0$ on an $L\times M$ rectangular lattice we find
that when all the hopping integrals are positive, the conductance is
independent of $L/M$, but if the distribution includes both the
positive and negative hopping integrals, the conductivity becomes
shape independent. Diagonal disorder corresponds to atomic
scatterers such as doped impurities, while OD disorder without sign
changes of hopping integrals represents randomness of bond lengths
caused by topological corrugations or warping in the sheets. For
graphene OD disorder is more important than diagonal disorder owing
to the ultrathin one-layer structure and the purity requirement in
preparing the samples. In both types short-range disorder can exist
and the intervalley scattering is not excluded. Thus the obtained
results should be different from those by using models without
intervalley scattering due to different symmetry classes. The sign
randomness of hopping integrals can be produced by geometric or
Berry phase $\sim \pm \pi$ which is acquired by electron
wavefunctions in every period of cyclic evolution around Dirac point
in parameter space. The Dirac point can serve as the degenerate
point for geometric phase in ``slow" cyclic evolutions, such as
in-plane vibrations of displacement between two sublattices, due to
the specific dispersion relation and electron-lattice interaction.

\section{Basic model and formalism}

The main features of the band structure in graphene can be well
described by a tight-binding Hamiltonian with one $\pi$ orbital per
site on a honeycomb lattice \cite{8a}:
\begin{equation}
  H= \sum_n \epsilon_n a^\dag_n a_n\sum_{\langle n,n' \rangle}
  +t_{nn'} (a^\dag_n a_{n'} +a^\dag_{n'} a_n),
  \end{equation}
where $a^\dag_n$ ($a_n$) creates (annihilates) an electron at site
$n$, $\epsilon_n$ is the energy level at site $n$, $\langle \ldots
\rangle$ denotes the nearest-neighbor (NN) sites, and $t_{nn'}$ is
hopping integral between NN sites $n$ and $n'$. Here, the spin
indices are not explicitly included. In the absence of disorder,
$\epsilon_n=0$ for all $n$ and $t_{nn'}=t_0$ for all $\langle nn'
\rangle$, this Hamiltonian leads to the 4-component Dirac fermion
dispersion relation near the Dirac point $E=0$. The diagonal and
off-diagonal disorder can be introduced by adopting random variables
$\epsilon_n$ and $t_{nn'}$ satisfying distribution probabilities
$P_d (\epsilon_n)$ and $P_o (t_{nn'})$, respectively. These
probability functions define strengths and types of disorder in the
system. Diagonal disorder represents potential fluctuations due to
impurities or due to randomness on substrate surface. A most
commonly used distribution probability for diagonal disorder is the
square function
\begin{equation}
   P_d (\epsilon_n) =\left\{ \begin{array}{l} 1/W, \text{  for   }
    -W/2 \leq \epsilon_{n} \leq W/2, \\ 0, \text{   otherwise},
    \end{array} \right.
    \end{equation}
where $W$ is the distribution width describing the strength of
disorder. For OD disorder, there are different distribution
functions describing different types of randomness. If we consider
slight fluctuations of bond lengths around their average value due
to lattice distortions, the following distribution is suitable
\begin{equation}
  \label{p01}
    P_{o1} (t_{nn'}) = \left\{ \begin{array}{l} 1/\lambda , \text{  for   }
    t_0-\lambda /2 \leq t_{nn'} \leq t_0 +\lambda /2, \\ 0, \text{   otherwise},
    \end{array} \right.
    \end{equation}
where $\lambda$ ($ < 2|t_0|$) is a measure of the bond-length
fluctuations which make the NN hopping integrals randomly shift from
$t_0$.

In $P_{o1}(t_{nn'})$ all hopping integrals have the same sign as
$t_0$, reflecting that only changes of magnitude of $t_{nn'}$ are
considered. In some cases, however, $t_{nn'}$ not only fluctuate in
magnitude, but also change in sign due to specific physical
mechanisms. One of these possible mechanisms is the geometric phase
due to dynamical vibrations. The on-plane displacement between two
sublattices, denoted as ${\bf d}=(d_x,d_y)$, changes all the NN bond
lengths and thus linearly couples with electron states in the regime
of small $|{\bf d}|$. At a Dirac point ($E=0$) the states are doubly
degenerate at ${\bf d}=0$. The displacement ${\bf d} \neq 0$
linearly removes the degeneracy, leading to conical energy surfaces
in the parameter space spanned by $d_x$ and $d_y$ with the
degenerate point situated at the origin $d_x=d_y=0$. Owing to
thermal excitations or even to zero-point oscillations, $d_x$ and
$d_y$ vibrate and both of them are dependent on time with period
$T=2\pi/\omega$ where $\omega$ is the frequency of vibrations.
Usually the amplitudes and relative phase of $d_x$ and $d_y$
vibrations are statistically random. For generic values of them, the
track in the $d_x$-$d_y$ parameter space in a period is an ellipse
encircling the degenerate point. Thus, according to the Berry
theorem, in each period of vibrations electron wavefunctions at
Dirac point will acquire a geometric phase $\sim \pm \pi$ in
addition to the ordinary dynamical phase. At the same time in this
period the tunneling electron has traversed a path of length $v_F T$
which equals $v_F T/l_0$ NN bonds with $v_F$ and $l_0$ being the
Fermi velocity and bond length respectively. This means that the
electron states change sign in traversing every $v_F T/l_0$ bonds.
Such an effect occurs very naturally since the vibrations and
electron transport are proceeding simultaneously. To include such an
effect in a primary way, we introduce the sign-changing probability
of an NN hopping $s$. This probability can be estimated from the
length by which the electron state changes sign due to the geometric
phase, i.e., $s \sim l_0/(v_F T)$. With such a sign-change model the
wavefunction can automatically change sign in traversing every
length of $v_F T$ which, averagely, contains only one hopping with
opposite sign. This is a special gauge in which the geometric phase
acquired in a path of $v_F T/l_0$ bonds is concentrated into one
bond in this path. This gauge is valid under the condition that the
interferences between paths shorter than $v_F T$ are not relevant.
As we will see below, the transport properties at Dirac point in the
case of OD disorder (including the sign changes) are
scale-invariant. So the short-range processes are not important and
the above condition can be satisfied. This sign-change mechanism is
different from the dephasing effect in the model of Ref.
\onlinecite{morozov} where the distortions are fluctuating in the
space (the ripples) but are static in time. We introduce the
following distribution of $t_{nn'}$
\begin{equation}
 \label{signn}
   P_o( t_{nn'}) = s P_{o1}(t_{nn'}) + (1-s) P_{o1}( -t_{nn'}),
   \end{equation}
to account for the sign-change probability $s$. When $s=0$ or $s=1$,
all the hopping integrals have the same sign and $P_o(t_{nn'})$
reduces to $P_{o1}(t_{nn'})$. At $s=0.5$ the probability of the sign
changes is maximal. Below we set $t_0$ as energy units.

\begin{figure}[htb]
\includegraphics[width=16.8cm]{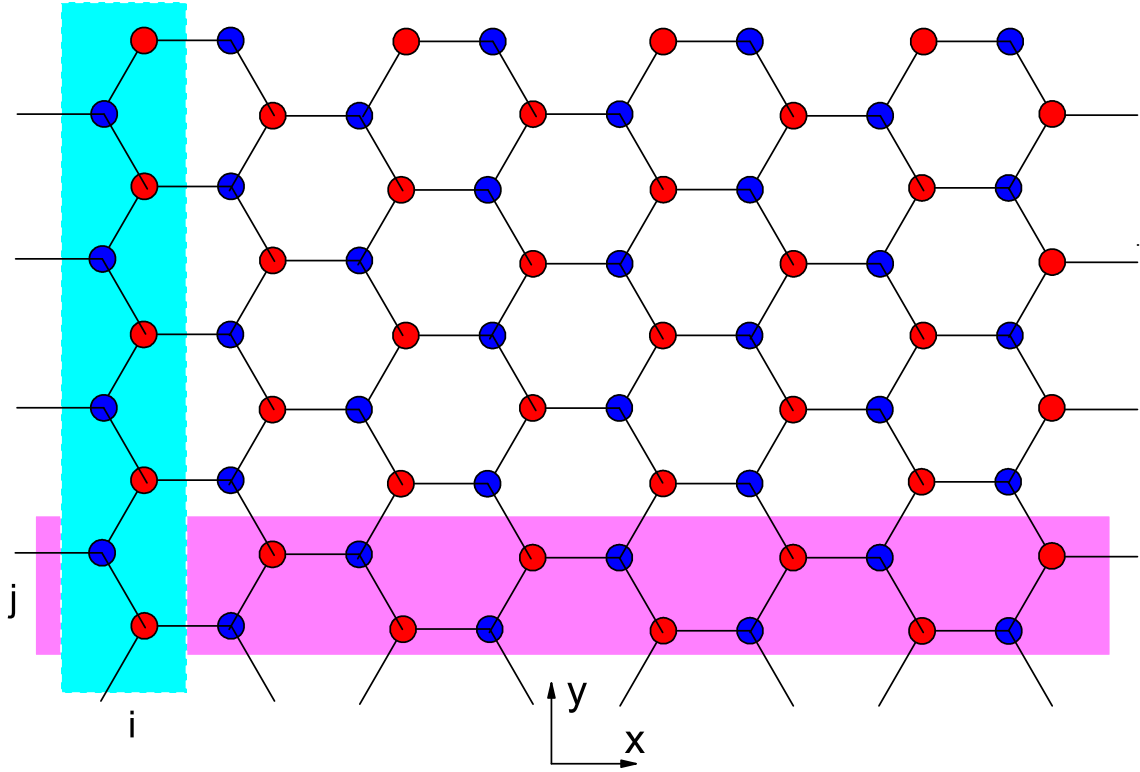}
\caption{Structure of graphene lattice. $i$ and $j$ are indices of
columns (shaded stripe along the $y$ direction) and rows (shaded
stripe along the $x$ direction), and the overlapping area of two
stripes is a basic cell where the lower and upper sites correspond
to sublattices $l=1$ and $l=-1$, respectively. }
\end{figure}

The wavefunction of a fermion can be written as
\begin{equation}
  \psi = \sum_n c_n | n\rangle,
\end{equation}
where $c_n$ is the amplitude on site $n$. At a given energy $E$ the
amplitudes are given by Schr\"{o}dinger equation $H \psi =E \psi$.
The propagation properties of an electron on a finite lattice can be
given by a relation between amplitudes on both sides. This is
usually formulated by the transfer-matrix method or by the Green
function. We note that graphene lattice can be divided into two
sublattices as distinguished by red and blue sites in Fig. 1. From
the tight-binding nature of the Hamiltonian and the structure shown
in Fig. 1 it can be seen that for given $E$ all amplitudes can be
iteratively calculated along the $x$ ($y$) direction if the
amplitudes in the light-cyan shaded column (light-magenta shaded
row) are given. From this a transfer matrix along the $x$ or $y$
direction can be given. We identify a site $n$ with three integers,
$n \equiv (i,j,l)$, where $i$ and $j$ index the column and row,
respectively, and $l=\pm 1$ specifies the sublattice which the site
$n$ belongs to. In this notation the iterative relation for
amplitudes along the $x$ direction can be derived from
Schr\"{o}dinger equation as
\begin{equation}
 \label{11}
  c_{i+1,j,-1}
  =\frac{ (E-\epsilon_{i,j,1}) c_{i,j,1} -
  t_{i,j,1;i,j,-1}
  c_{i,j,-1} - t_{i,j,1; i,j+(-1)^i,-1} c_{i,j+(-1)^i,-1}}{t_{i,j,1;i+1,j,-1}},
  \end{equation}
\begin{equation}
  \label{12}
  c_{i+1,j,1} =\frac{(E-\epsilon_{i+1,j,-1}) c_{i+1,j,-1} - t_{i+1,j,-1;i,j,1}
  c_{i,j,1} - t_{i+1,j,-1; i+1,j+(-1)^i,1} c_{i+1,j+(-1)^i,1}}{t_{i+1,j,-1;i+1,j,1}},
  \end{equation}
For a system with finite width in the $y$ direction, one can define
vectors for columns by ${\bf u}_{i,l}\equiv ( c_{i,1,l},c_{i,2,l},
\ldots, c_{i,M,l})^T$, where $M$ is the number of cells in a column.
Then Eqs. (\ref{11}) and (\ref{12}) can be written in a matrix form:
\begin{equation}
  {\bf u}_{i+1,-1} = \hat{T}_{1,i} {\bf u}_{i,1} +\hat{T}_{2,i}{\bf
  u}_{i,-1},
  \end{equation}
\begin{equation}
  {\bf u}_{i,1} = \hat{T}_{3,i} {\bf u}_{i+1,1} +\hat{T}_{4,i}{\bf
  u}_{i+1,-1},
  \end{equation}
where elements of matrices are given by
\begin{equation}
  \{\hat{T}_{1,i}\}_{ j,j'} = \frac{E-
  \epsilon_{i,j,1}}{t_{i,j,1;i+1,j,-1}} \delta_{j,j'} ,
  \end{equation}
\begin{equation}
  \{\hat{T}_{2,i}\}_{ j,j'} = -\frac{
  t_{i,j,1;i,j,-1}}{t_{i,j,1;i+1,j,-1}} \delta_{j,j'} -\frac{
  t_{i,j,1;i,j+(-1)^i,-1}}{t_{i,j,1;i+1,j,-1}} \delta_{j+(-1)^i,j'},
  \end{equation}
\begin{equation}
  \{\hat{T}_{3,i}\}_{ j,j'} = -\frac{
  t_{i+1,j,-1;i+1,j,1}}{t_{i+1,j,-1;i,j,1}} \delta_{j,j'} -\frac{
  t_{i+1,j,-1;i+1,j+(-1)^i,1}}{t_{i+1,j,-1;i,j,1}} \delta_{j+(-1)^i,j'},
  \end{equation}
\begin{equation}
  \{\hat{T}_{4,i}\}_{ j,j'} = \frac{E-
  \epsilon_{i+1,j,-1}}{t_{i+1,j,-1;i,j,1}} \delta_{j,j'} .
  \end{equation}
Here, one may use the periodic or open boundary condition at the
ends $j=1$ and $j=M$. We have the iterative relation for whole
column
\begin{equation}
 \left( \begin{array}{l} {\bf u}_{i+1,-1} \\ {\bf u}_{i+1,1} \end{array}
 \right) = \hat{T}_i  \left( \begin{array}{l} {\bf u}_{i,-1} \\ {\bf u}_{i,1} \end{array}
 \right),
 \end{equation}
where the transfer matrix is
 \begin{equation}
  \hat{T}_i = \left( \begin{array}{ccc} \hat{T}_{2,i}, &  & \hat{T}_{1,i}
  \\ -\hat{T}_{3,i}^{-1} \hat{T}_{4,i} \hat{T}_{2,i}, &   & \hat{T}_{3,i}^{-1}-\hat{T}_{3,i}^{-1}
  \hat{T}_{4,i} \hat{T}_{1,i} \end{array} \right).
  \end{equation}
If there are $L$ columns in the $x$ direction, the total transfer
matrix which gives the relation between amplitudes at two end
columns is
\begin{equation}
   \hat{T} = \prod_{i=1}^L \hat{T}_{L-i+1}.
   \end{equation}

If we want to investigate the propagation properties in the $y$
direction, we have to establish the transfer matrix for rows. The
relation between two adjacent rows is written as
\begin{equation}
 \left( \begin{array}{l} {\bf v}_{j+1,-1} \\ {\bf v}_{j+1,1} \end{array}
 \right) = \hat{V}_j  \left( \begin{array}{l} {\bf v}_{j,-1} \\ {\bf v}_{j,1} \end{array}
 \right),
 \end{equation}
where ${\bf v}_{j,l}$'s are vectors denoting rows with length $M$,
${\bf v}_{j,l}\equiv ( c_{1,j,l},c_{2,j,l}, \ldots, c_{M,j,l})^T$,
and the elements of the transfer matrix are
 \[
 \{ \hat{V}_j \}_{i,l;i',l'}= \delta_{l,-(-1)^i}
 \left\{ C^{(1)}_{i,j}\delta_{-l,l'}
 \delta_{i,i'} - C^{(2)}_{i,j}
\delta_{l,l'}\delta_{i,i'}
  -C^{(3)}_{i,j}
\delta_{l,l'} \delta_{i+(-1)^i,i'} \right. \]
\[
  + C^{(4)}_{i,j}
(C^{(1)}_{i,j}\delta_{l,l'}
 \delta_{i,i'} - C^{(2)}_{i,j}
\delta_{l,-l'}\delta_{i,i'}
  -C^{(3)}_{i,j}
\delta_{l,-l'}\delta_{i+(-1)^i,i'}) -C^{(5)}
\delta_{l,l'}\delta_{l,-(-1)^i}\delta_{i,i'}
\]
\begin{equation}
 \left.
  -C^{(6)} (C^{(1)}_{i-(-1)^i,j}\delta_{l,-l'}
 \delta_{i-(-1)^i,i'} - C^{(2)}_{i-(-1)^i,j}
\delta_{l,l'}\delta_{i-(-1)^i,i'}
  -C^{(3)}_{i-(-1)^i,j}
\delta_{l,l'}\delta_{i-(-1)^i\cdot 2,i'})
   \right\} ,
\end{equation}
where
\[
 C^{(1)}_{i,j}= \frac{E- \epsilon_{i,j,(-1)^i}}{t_{i,j,(-1)^i;
 i,j+1,-(-1)^i}},
 \,\,\,
 C^{(2)}_{i,j}=\frac{
t_{i,j,(-1)^i;i,j,-(-1)^i}}{t_{i,j,(-1)^i; i,j+1,-(-1)^i}}, \,\,\,
 C^{(3)}_{i,j} =\frac{
t_{i,j,(-1)^i;i+(-1)^i,j,-(-1)^i}}{t_{i,j,(-1)^i; i,j+1,-(-1)^i}},
\]
\[
  C^{(4)}_{i,j}=\frac{E-
\epsilon_{i,j+1,-(-1)^i}}{t_{i,j+1,-(-1)^i; i,j+1,(-1)^i}}, \,\,\,
  C^{(5)}_{i,j}=\frac{
t_{i,j+1,-(-1)^i; i,j,(-1)^i}}{t_{i,j+1,-(-1)^i; i,j+1,(-1)^i}},
\,\,\,
  C^{(6)}_{i,j}=\frac{
t_{i,j+1,-(-1)^i; i-(-1)^i,j+1,(-1)^i}}{t_{i,j+1,-(-1)^i;
i,j+1,(-1)^i}}.
\]
The total transfer matrix for $L$ rows is
\begin{equation}
   \hat{V} = \prod_{i=1}^L \hat{V}_{L-i+1}.
   \end{equation}

For a finite $M\times L$ system, there are $M$ propagating channels
along the length $L$. The Hermitian matrix, $ (\hat{T}^{\dag
}\hat{T})^{1/2L} $ or $ (\hat{V}^\dag \hat{V})^{1/2L}$, has $2M$
eigenvalues. These $2M$ eigenvalues should be positive and come in
inverse pairs due to the unitarity of the matrix
\cite{g6,g20,g21,g22}. The $M$ positive logarithms of eigenvalues,
denoted as $\gamma_{i}$ in an ascending order with index $i$, are
Lyapunov exponents (LE) reflecting the exponential decay of the
corresponding channels. Obviously, the logarithms of other $M$
eigenvalues are the negatives of $\gamma _{i}$. The LEs will
approach $0$ when the disorder vanishes. The localization length is
defined as the reciprocal of the smallest LE $\xi \equiv 1/\gamma
_{1}$. The localization length plays a crucial role in the
localization theory.

The zero-temperature conductance $g$ of a rectangular $M\times L$
system in units of $2e^{2}/h$, with two spins being taken into
account, can be evaluated using $M$-channel Landauer formula
\cite{g24,g25,g26}
\begin{equation}
g=\text{Tr}(\hat{t}^{\dagger }\hat{t}),  \label{6}
\end{equation}%
where the $M\times M$ transmission matrix $\hat{t}$ describes the
transmission of electrons from one lead to the other in the
longitudinal direction. Here we assume that the leads are connected
to all the $M$ channels and they are perfect metals with band width
much larger than that of graphene. Formula (\ref{6}) can be
expressed with the LEs as \cite{g6,g22}
\begin{equation}
 \label{cond}
g=\sum_{i=1}^{M}\frac{1}{\cosh^2 (\gamma_{i}L)}.
\end{equation}%

Numerically, LEs can be calculated by using the standard method of
Gram-Schmidt re-orthonormalization after each a few, say, ten steps
of multiplication of transfer matrices \cite{g3,g4}. This is
equivalent to the diagonalization of $\hat{T}^\dag \hat{T}$ or
$\hat{V}^\dag \hat{V}$, but avoids terrible overflow and loss of
precision on the computer. On the other hand, although the
calculations of transfer matrices along the $x$ and $y$ directions
are certainly different, there is no essential difference in the
calculated results between these two directions. This is owing to
the fact that the dispersion relation is conic around the Dirac
point which is isotropic in the $x$ and $y$ directions. Below we
will only present the results obtained for the transmission along
the $y$ direction.

\section{Localization in system with diagonal disorder}

First we investigate the case of diagonal disorder, i.e., $W\neq 0$,
$\lambda =0$, and $s=0$. The lowest LE $\gamma_1(M)$ is calculated
for very long ($L \sim 10^6$) strip of width $M$. The rescaled
localization length is defined as $
\frac{\Lambda(M)}{M}=\frac{1}{\gamma_1(M)M} $. For given energy and
disorder strength, the localization behavior of the fermions can be
determined by the scaling dependence of $\frac{\Lambda(M)}{M}$: If
$\frac{\Lambda(M)}{M}$ increases with $M$, the states are extended.
For localized states $\frac{\Lambda(M)}{M}$ is decreased by
increasing $M$. The $M$-independence of $\frac{\Lambda(M)}{M}$
corresponds to the fixed point separating regions of extended and
localized states. In Fig. 2(a), we plot the $M$ dependence of
$\frac{\Lambda(M)}{M}$ for states at $E=0$ in systems with different
strengths of diagonal disorder. For all investigated strengths of
diagonal disorder, the rescaled localization length always decreases
with increasing width. The large fluctuations in the curve of $W=1$
are due to numerical errors in the Gram-Schmidt
re-orthonormalization procedure which become serious in the case
with $\frac{\Lambda(M)}{M}\gg 1$. This means that the states at the
Dirac point are localized by tiny diagonal disorder. In Fig. 2(b),
we plot the $M$-dependence of $\frac{\Lambda(M)}{M}$ for a given
strength of diagonal disorder but different energies. Although all
states are localized, the scaling behavior is much different between
the states at Dirac point and the other states: the decreasing slope
of the former is much larger than that of the latter. This implies
that the states at Dirac point are more easily localized than other
states by introducing the diagonal disorder in the thermodynamical
limit, in spite of their larger localization length in smaller
systems. This may be due to the small density of states at $E=0$ and
the breaking of the chiral symmetry by diagonal disorder. The
localization behavior of the states is consistent with the common
theories of the Anderson localization in 2D.

\begin{figure}[htb]
\includegraphics[width=8.8cm]{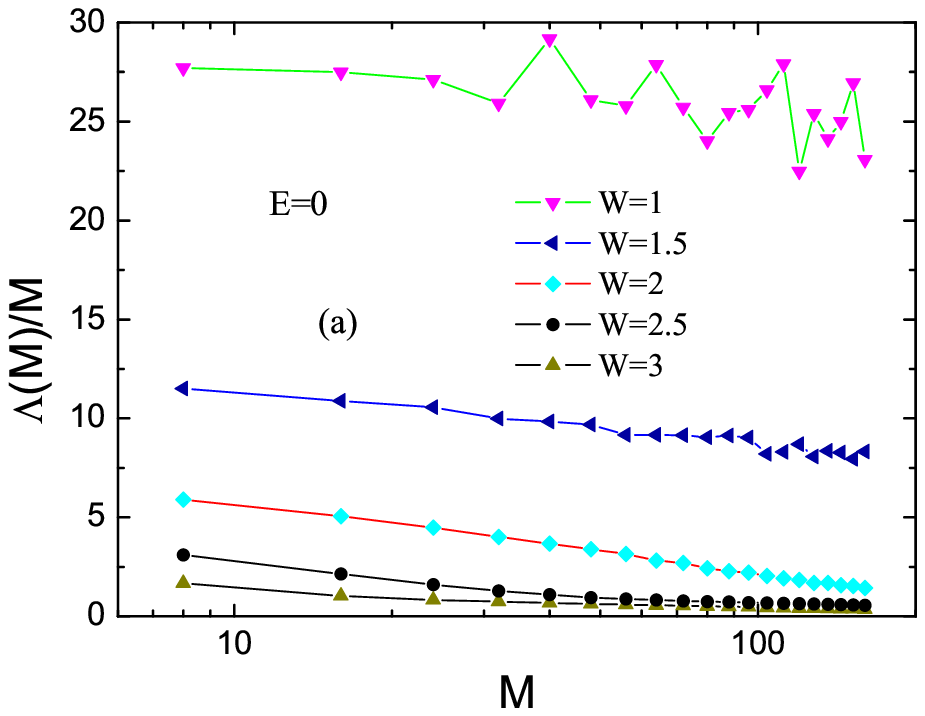}
\includegraphics[width=8.8cm]{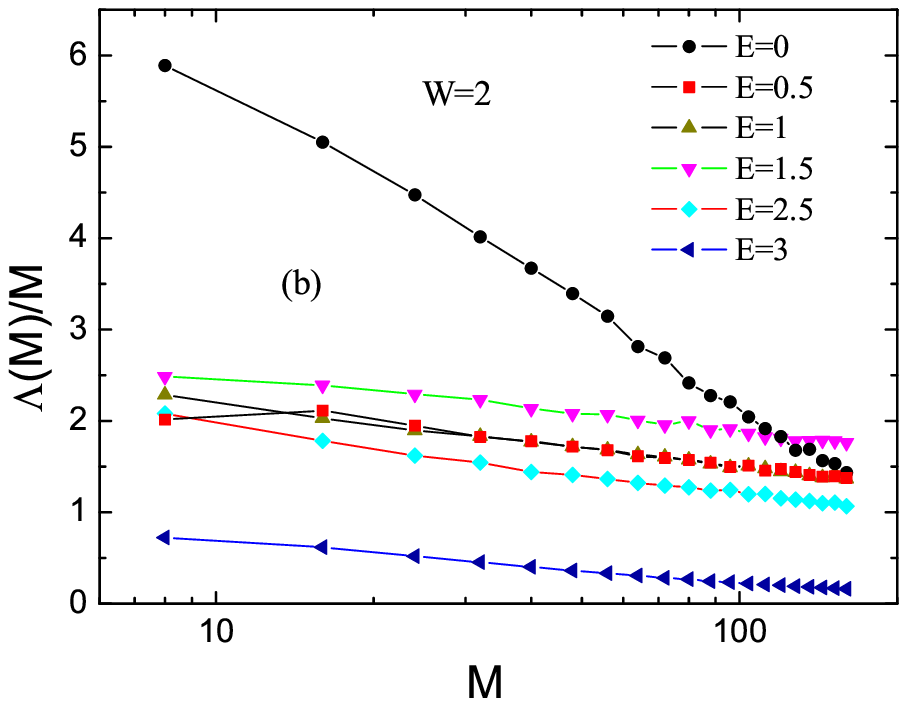}
\caption{(Color online) Scaling behavior of rescaled localization
length in the presence of diagonal disorder. (a)
$\frac{\Lambda(M)}{M}$ versus $M$ at the Dirac point ($E=0$) for
different strengths of disorder. (b) $\frac{\Lambda(M)}{M}$ versus
$M$ for a given strength of disorder and various energies.  }
\end{figure}

\section{Delocalization at Dirac point in the case of off-diagonal disorder
without sign changes}

Differently from diagonal disorder, off-diagonal disorder does not
break the chiral symmetry at Dirac point, leading to an essential
difference in the localization. The scaling analysis of the rescaled
localization length and the conductance at Dirac point is shown in
Fig. 3 by using the distribution function $P_{o1}$ in Eq.
(\ref{p01}). This corresponds to the case of randomly warped or
corrugated lattice where bond lengths are randomly shifted from the
perfect value. From Fig. 3(a) we can see that the rescaled
localization length is independent of $M$ for all disorder
strengths. This implies that the localization length at the
thermodynamical limit is infinite, corresponding to delocalized
states at $E=0$. The conductance of an $M\times M$ system calculated
from Eq. (\ref{cond}) is also independent of $M$ for all
investigated strengths of off-diagonal disorder [see Fig. 3(b)]. So
the conductance at the thermodynamical limit can be determined from
these size independent values. The dependence of $g$ on the disorder
strength is shown in the inset of Fig. 3(b). For relatively weak
off-diagonal disorder ($\lambda < 1$) the conductance at Dirac point
is about $\frac{4e^2}{h}$, corresponding to the quantum conductance
of 4 channels (2 spin channels and 2 equivalent channels from the
chiral symmetry at $E=0$). In fact, in a system with off-diagonal
disorder, the calculated Lyapunov exponents at $E=0$ can be grouped
into pairs, i.e., $\gamma_{2i-1}= \gamma_{2i}$ for integer $i$ in
the range between 1 and $\frac{M}{2}$, corresponding to two
equivalent channels with the chiral transformation. These channels
have the localization length proportional to the width $M$. At the
thermodynamical limit they become infinite. By increasing the
disorder strength from $\lambda =1$ towards its maximum value
$\lambda=2$, these channels are still delocalized at the
thermodynamical limit but the conductance rapidly drops, as can be
seen from the inset of Fig. 3(b).

\begin{figure}[htb]
\includegraphics[width=8.8cm]{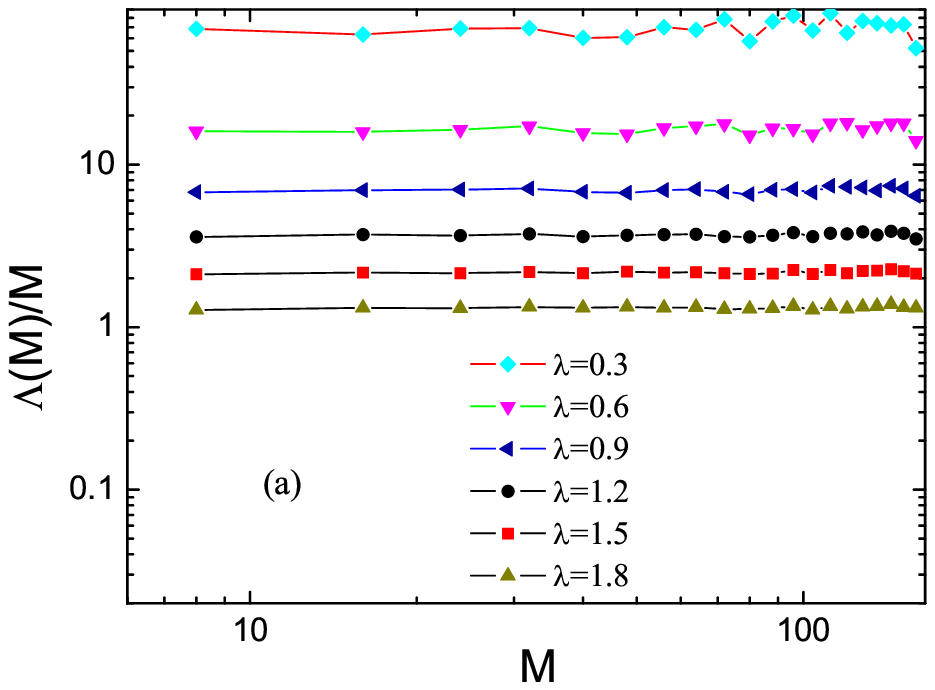}
\includegraphics[width=8.8cm]{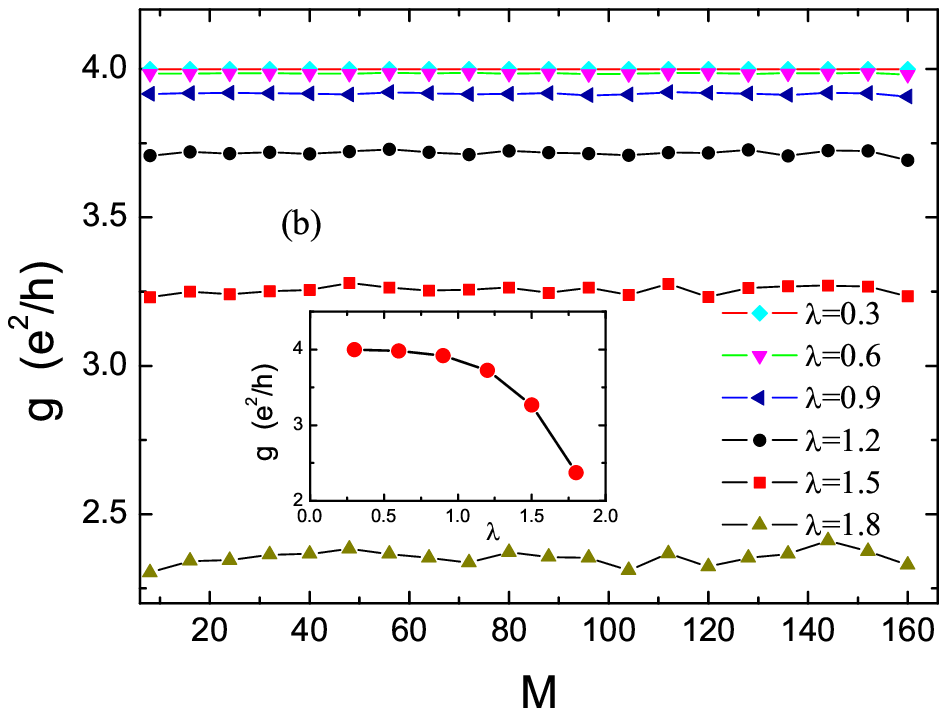}
\caption{(Color online) Scaling behavior of the rescaled
localization length (a) and the conductance of an $M\times M$ system
(b) at the Dirac point ($E=0$) in the presence of off-diagonal
disorder. The inset of (b) shows the dependence of $g$ on the
strength of off-diagonal disorder $\lambda$.  }
\end{figure}

The delocalization behavior at $E=0$ in graphene with the
off-diagonal disorder is not consistent with the usual localization
behavior of 2D disordered systems. One may attribute this
inconsistency to the chiral symmetry at $E=0$ which is conserved in
off-diagonal disorder but destroyed in diagonal disorder. So the
localization at energies $E\neq 0$, where the chiral symmetry no
longer exists, could be expected. In Fig. 4, we plot the scaling
behavior of the rescaled localization length and the conductance of
an $M\times M$ system for different energies. Except for the curves
of $E=0$, both the rescaled localization length and the conductance
decrease by increasing $M$ in the general trend if fluctuations of
some curves are ignored, suggesting the usual localization behavior
at $E\neq 0$. From Fig. 4(a), the rescaled localization length is
generally decreased by increasing $E$ from $E=0$ for a fixed $M$.
This means that in off-diagonal disorder the states of energies
farther apart from Dirac point are more localized. From Fig. 4(b),
however, for a given value of $M$ the conductance at larger $|E|$
has larger values. This can be explained by the fact that there are
more electron channels at larger $|E|$ in graphene which can
contribute to the conductance if $M$ is not very large.

\begin{figure}[htb]
\includegraphics[width=8.8cm]{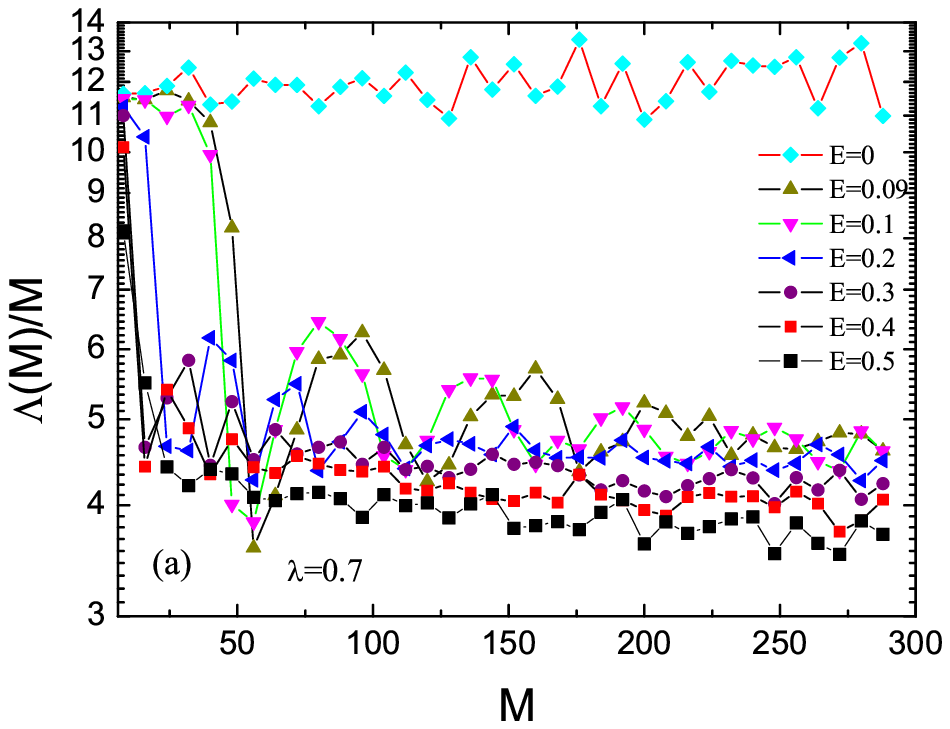}
\includegraphics[width=8.8cm]{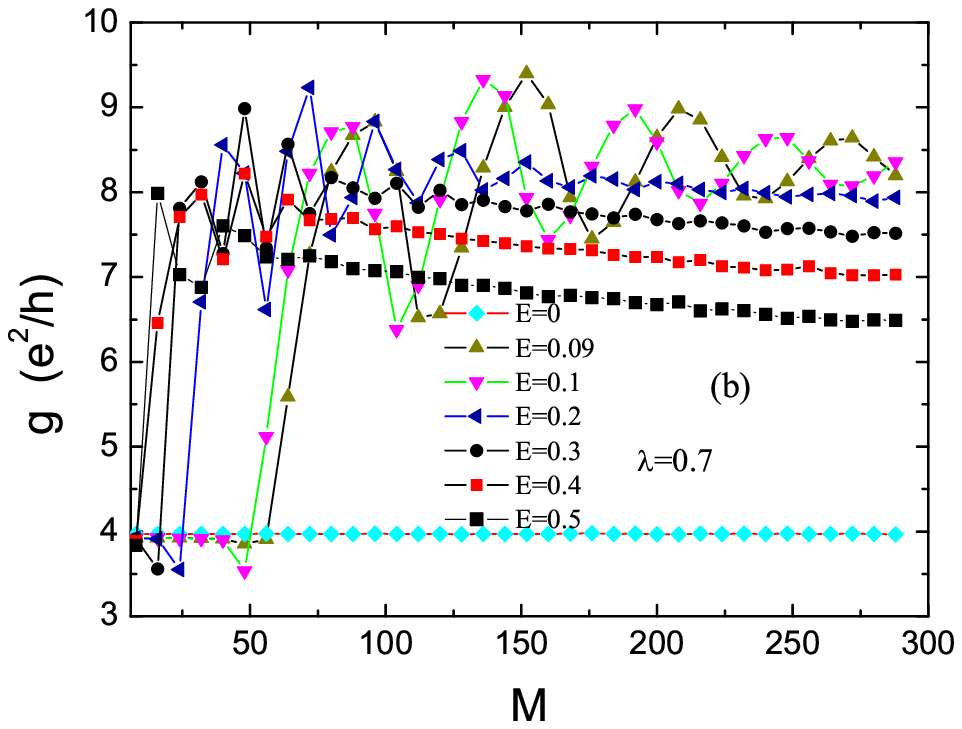}
\caption{(Color online) Scaling behavior of the rescaled
localization length (a) and the conductance of an $M\times M$ system
(b) for different energies in the presence of off-diagonal disorder.
}
\end{figure}

By using a scaling transformation $M \rightarrow M/\xi$ with
appropriate value of scaling factor $\xi$ for every energy $E$, the
curves in Fig. 4(a) for different energies can merge together. In
Fig. 5 we plot the common dependence of $\frac{\Lambda(M)}{M}$ on
$M/\xi$ for various energies. The scaling factor $\xi$ as a function
of $E$ is shown in the inset which can be fitted by a simple
function $\xi = \frac{a}{|E|}$ with prefactor $a= 2.66$. Since $\xi$
is just the localization length at the thermodynamical limit, this
scaling behavior further ensures the delocalization at $E=0$ and
describes how the localization length diverges when $E\rightarrow
0$. From the Dirac dispersion relation the wave length of electrons
is also inversely proportional to $|E|$. This suggests that in the
case of off-diagonal disorder the localization length and the
electron wave length have the same $E$ dependence and can be
regarded as the same length scale. We also notice that although the
states at $E=0$ are delocalized, there is no real metallic region
with finite range of $E$. This is certainly different from the
results in Refs.
\onlinecite{Ostrovsky,titov,Bardarson,nomura2,San-Jose} where the
disorder models free of intervalley scattering are adopted.

\begin{figure}[htb]
\includegraphics[width=8.8cm]{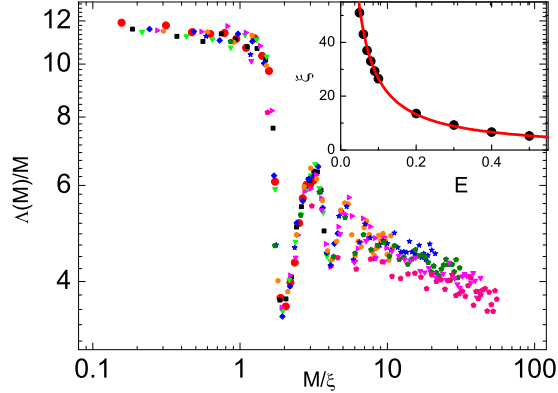}
\caption{(Color online) Common dependence of the rescaled
localization length $\frac{\Lambda}{M}$ on the scaled width
$\frac{M}{\xi}$ for various energies. Inset: $\xi$ as a function of
$E$ (circle symbols) which is fitted by function $\xi
=\frac{2.66}{E}$ (red curve).  }
\end{figure}

The different localization behaviors between $E=0$ and $E\neq 0$ can
be more clearly depicted by the spatial distributions of
wavefunctions as shown in Fig. 6. As can be seen from Fig. 6(a), the
delocalized state at $E=0$ is more likely to be a Cantor-set like
critical state, rather than a real metallic state. Such states may
provide propagation paths for current from one edge to the other. On
the other hand, the state at $E=1.8$ shown in Fig. 6(b) has
significant amplitudes only in some isolated islands, corresponding
to typical localization.

\begin{figure}[htb]
\includegraphics[width=8.8cm]{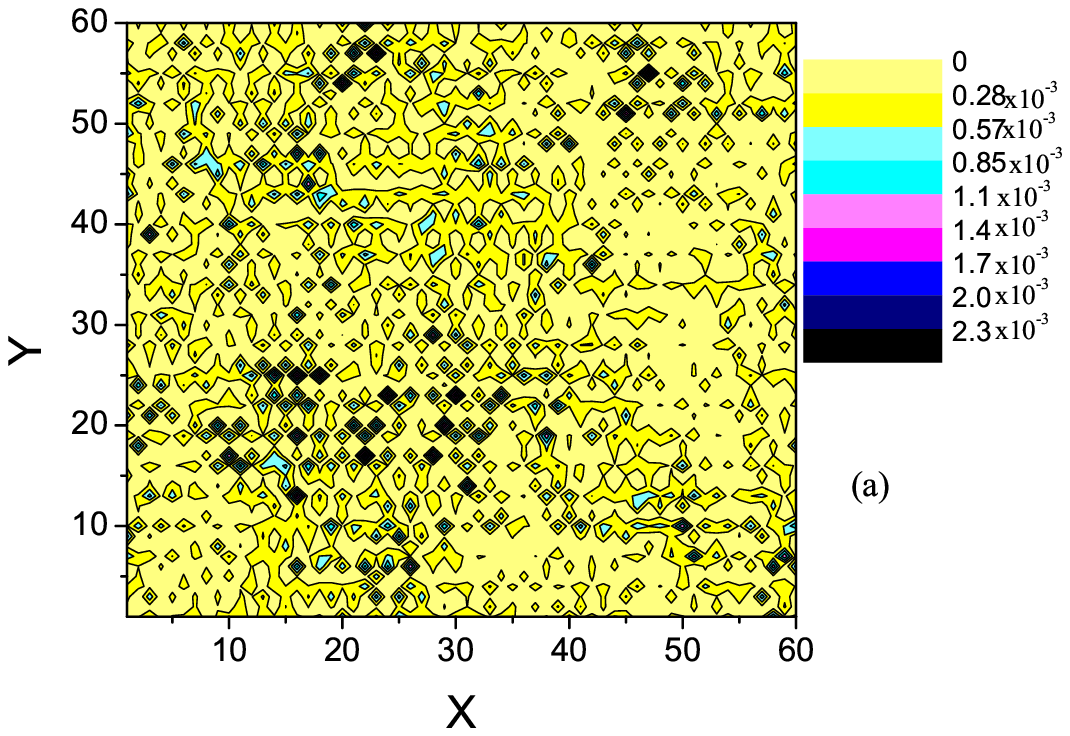}
\includegraphics[width=8.8cm]{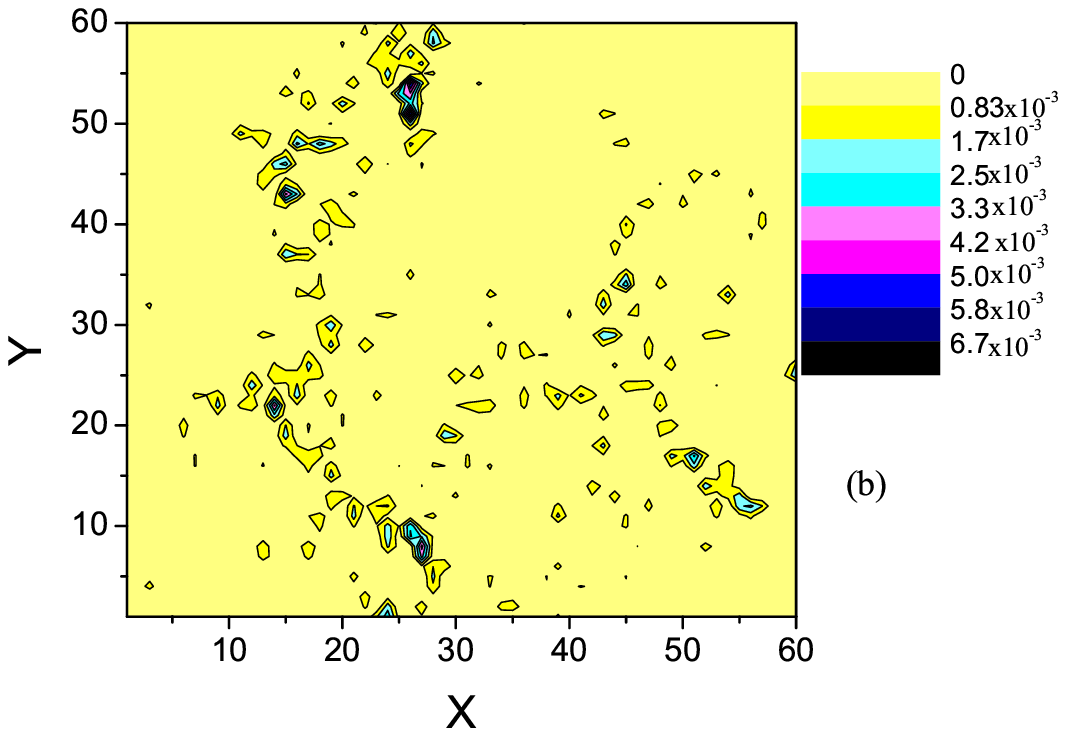}
\caption{(Color online) Contour plot of spatial distributions of
wavefunctions at (a) $E=0$ and (b) $E=1.8$ in a $60 \times 60$
system in off-diagonal disorder of strength $\lambda =0.7$. }
\end{figure}

In the case of off-diagonal disorder there are 4 delocalized
channels at Dirac point, giving conductance $\frac{4e^2}{h}$. This
leads to almost shape-independent conductance and shape-dependent
conductivity. The situation is similar to the case of perfect
lattice investigated by Tworzyd\/{l}o {\it et al}. in Ref.
\onlinecite{twor}. Following the procedure proposed in Ref.
\onlinecite{twor}, we calculate the conductivity from the
conductance of a $M\times L$ rectangular sample. The obtained shape
dependence of the conductivity is shown in Fig. 7. This behavior is
consistent with the nature of the quantum transport in 2D, but is
inconsistent with the experiments where the conductivity is shape
independent. This means that to explain this experimental result we
have to invoke some mechanism which could convert the quantum
transport to the classical-like transport.

\begin{figure}[htb]
\includegraphics[width=8.8cm]{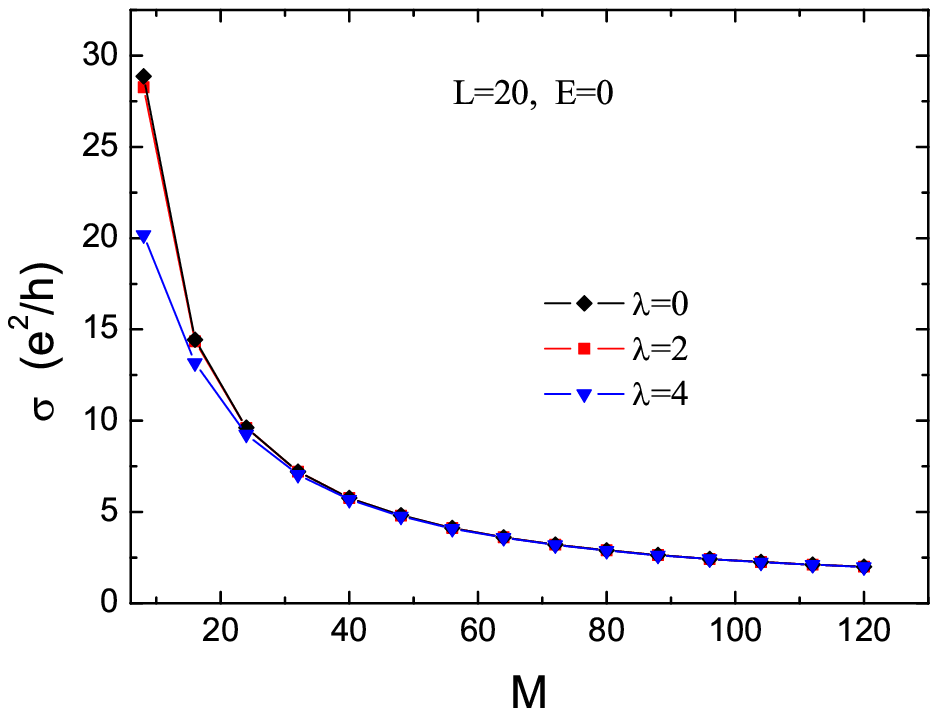}
\caption{(Color online) Width dependence of conductivity of an
$M\times L$ rectangular system with a given length at $E=0$ for
different strengths of off-diagonal disorder. }
\end{figure}

\section{Classical-like transport at Dirac point in the case of off-diagonal disorder
with sign changes}

The quantum transport processes could be drastically changed if the
phases of wavefunctions become uncertain. This may happen due to the
geometric phase which can be acquired by electrons during a cyclic
evolution of ``slow" degrees of freedom around a conic point. The
additional geometric phase of $\sim \pm \pi$ corresponds to sign
changes of the wavefunction during the motion of tunneling
electrons. In the diffusion regime, if the sign changes randomly
happen during the spatial motion, they can be modeled by a random
distribution of positive and negative signs of hopping integrals. In
probability $P_o(t_{nn'})$ of Eq. (\ref{signn}) we include both the
sign randomness and the magnitude randomness of the hopping
integrals which are specified by parameters $s$ and $\lambda$,
respectively.

First we set $\lambda =0$ and focus on the effect of sign
randomness. In Fig. 8 we plot the scaling behavior of the rescaled
localization length and the conductance of an $M\times M$ system for
different values of $s$. Similarly to the case of off-diagonal
disorder, the states at the Dirac point in the presence of the sign
randomness are also delocalized in the sense of finite-size scaling
analysis. Differently from the behavior in the off-diagonal disorder
shown in Fig. 3, effects of sign randomness on the rescaled
localization length and on the conductance are opposite:
$\frac{\Lambda (M)}{M}$ in $s=0.5$ is smaller than that in $s=0.2$,
but $g$ in $s=0.5$ is larger than that in $s=0.2$. This is strange
because $\frac{\Lambda(M)}{M}$ corresponds to the channel with the
largest localization length which usually provides the leading
contribution to the conductance. The only explanation for this is
that the other channels also give significant contributions in the
case of sign randomness. An expectable consequence of this is a new
shape dependence of the conductivity, since the number of the
tunneling channels is no longer restricted.

\begin{figure}[htb]
\includegraphics[width=8.8cm]{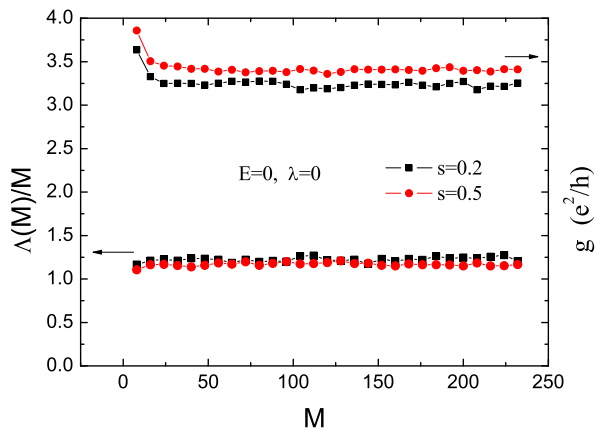}
\caption{(Color online) Scaling behavior of the rescaled
localization length and the conductance of an $M\times M$ system in
the presence of sign randomness of hopping integrals obeying
probability $P_o(t_{nn'})$ with different values of $s$. }
\end{figure}

The shape dependence of the conductivity can be investigated by
calculating the conductance of $M\times L$ rectangular samples. The
obtained conductivity as a function of ratio $M/L$ for various sizes
is shown in Fig. 9. Except for small values of $M/L$, the
conductivity has values near $\frac{4e^2}{h}$, independent of $M/L$
and $M$, in consistence with the experimental findings.
Surprisingly, this value is even larger than $\frac{4e^2}{\pi h}$
obtained in ballistic graphene \cite{twor,perf} in spite of the
disorder introduced by the sign randomness of hopping integrals. To
get insight into this anomalous feature, we adopt the average
spacing between successive pairs of Lyapunov exponents denoted as
$\delta\gamma$, i.e., $\gamma_{2i+1}-\gamma_{2i-1} \sim
\gamma_{2i+2}-\gamma_{2i} \sim \delta \gamma$. Notice that in this
case the Lyapunov exponents are grouped into pairs, corresponding to
two equivalent channels from the chiral symmetry. By keeping only
the leading exponential term of $\cosh^2 (\gamma_i L)$, Eq.
(\ref{cond}) can be approximately rewritten as
\begin{equation}
   g \approx \sum_{i=1}^{M/2} 8 e^{-2\gamma_{2i-1}L} \approx \frac{ 8 e^{-2\gamma_1
   L} (1- e^{-(M+2)L \delta \gamma})}{1- e^{-2L \delta \gamma}},
   \end{equation}
in units of $\frac{2e^2}{h}$. In a perfect system of width $M$ and
using the periodic boundary condition, the spacing $\delta \gamma$
for states at $E=0$ is $\delta \gamma = 2\pi /M$ due to the Dirac
fermion dispersion relation \cite{perf}. One may reasonably suppose
that $\delta \gamma \propto 1/M$ is still held in the case of OD
disorder. Then, for $M\gg L$, the conductivity of an $M\times L$
rectangular sample can be calculated as
\begin{equation}
  \label{ooo}
  \sigma = \frac{ g L}{ M} \approx \frac{4 e^{-2\gamma_1 L}
  }{\delta \gamma M}.
  \end{equation}
So for the perfect lattice $\sigma = \frac{2}{\pi}$ in units of
$\frac{2e^2}{h}$ \cite{twor,perf}. In the case of sign randomness,
the values of $M \delta \gamma$ are in the range $[0.9,2.5]$ as
shown in the inset of Fig. 9. Together with the factor
$e^{-2\gamma_1 L}$, the obtained conductivity may be $\pi$ times of
that of the perfect lattice. Thus, the $\pi$ factor difference in
the conductivity is not trivial and reflects an essential change of
the fermion properties due to the phase uncertainty.

\begin{figure}
\includegraphics[width=8.8cm]{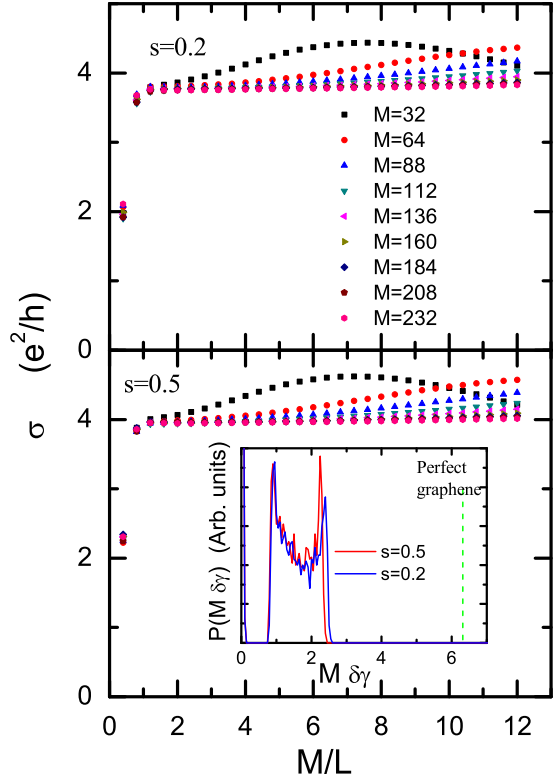}
\caption{(Color online) Calculated conductivity as a function of
ratio $M/L$ for rectangular sheets with different widths $M$. Inset
in the lower panel: Distribution function $P( M\delta \gamma)$ of
$M\delta \gamma$. The green dashed line indicates the
$\delta$-function distribution at $2\pi$ in the case of perfect
sheet. } \label{fig8}
\end{figure}

It is interesting to investigate how the localization behavior
changes in the presence of the magnitude randomness of hopping
integrals. The scaling behaviors of the rescaled localization length
and the conductance of an $M\times M$ system with both nonzero $s$
and $\lambda$ are shown in Fig. 10. It can be seen that the states
at $E=0$ are still delocalized and the opposite effects on the
rescaled localization length and on the conductance are more evident
than those shown in Fig. 8. In Fig. 11 we display the shape
dependence of the conductivity of the $M\times L$ rectangular
samples for $s=0.5$ and different nonzero values of $\lambda$.
Similarly to the case of $\lambda=0$, the conductivity is still
shape independent except for small values of $M/L$. This means that
the sign-randomness induced shape independence of the conductivity
is robust against the warping or corrugation disorder which causes
randomness of magnitudes of hopping integrals. It can also be seen
that the value of the conductivity is slightly reduced from
$\frac{4e^2}{h}$ by increasing the warping or corrugation disorder.
For reasonable strength of the warping disorder ($\lambda <1$),
however, the conductivity is above $\frac{3e^2}{h}$, in the range of
measured values in experiments.

\begin{figure}[htb]
\includegraphics[width=8.8cm]{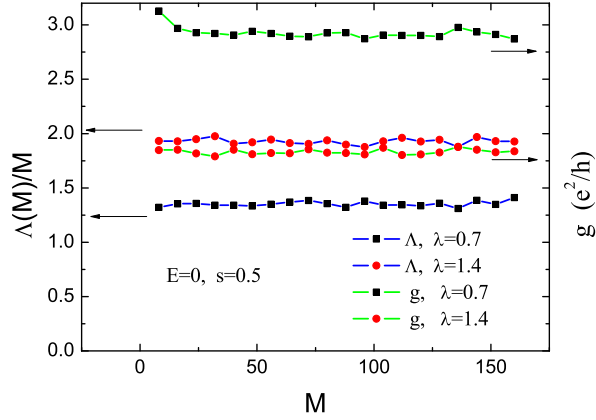}
\caption{(Color online) Scaling behavior of the rescaled
localization length and the conductance of an $M\times M$ system in
the presence of both sign randomness and magnitude randomness of
hopping integrals obeying probability $P_o(t_{nn'})$. }
\end{figure}

\begin{figure}
\includegraphics[width=8.8cm]{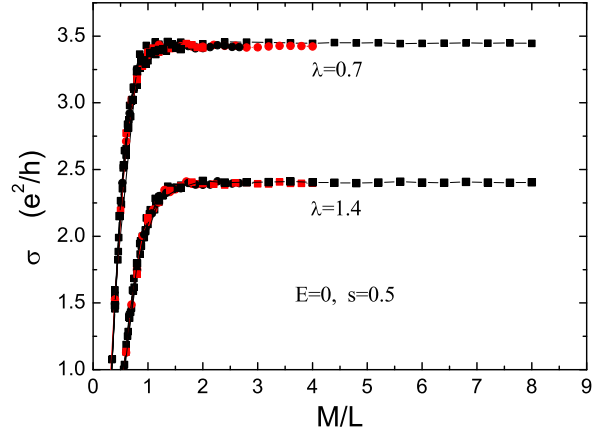}
\caption{(Color online) Calculated conductivity as a function of
ratio $W/L$ for rectangular sheets with different widths $M$ in the
presence of both sign randomness and magnitude randomness of hopping
integrals. }
\end{figure}

It is also interesting to investigate the scaling behavior of the
statistics of the minimum conductivity. In Fig. 12 we plot the
statistical distribution of the conductivity at $E=0$ for various
system sizes and for given values of $M/L$, $\lambda$, and $s$. We
note that the distribution of the conductivity is Gaussian-like, and
both the average and the distribution width are almost scaling
invariant. This is different from the statistics of a single
Lyapunov exponent in usual disordered system for which the
distribution width is $1/M$ dependent. This difference originates
from the fact that in the present system the conductivity is
contributed not only from the channel with the smallest LE but also
from other channels. As a consequence the statistics of the
conductivity given by Eq. (\ref{ooo}) is mainly determined by the
distribution of $\delta \gamma M$ rather than by the distribution of
a single LE. It can be easily verified that in the present case the
series of successional LEs times $M$, $\gamma_1M, \gamma_2M,\ldots,
$ are scaling invariant. Thus the statistics of $\delta \gamma M$ is
also scaling invariant, leading to the scaling invariant statistics
of the minimum conductivity.

\begin{figure}
\includegraphics[width=8.8cm]{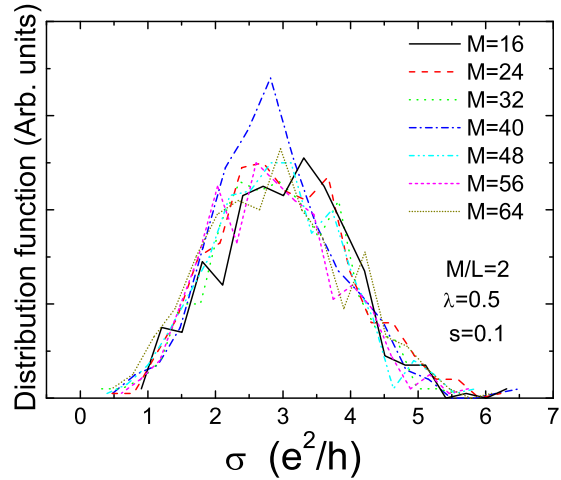}
\caption{(Color online) Statistical distribution of the conductivity
at $E=0$ for various system sizes in OD disorder with sign
randomness. The values of $M/L$, $\lambda$, and $s$ are given in the
figure. The statistics is taken from 500 random configurations. }
\end{figure}

\section{Conclusions}

We investigate the effect of different types of disorder on the
transport properties of electrons at Dirac point in graphene. From
the finite-size scaling analysis together with the transfer-matrix
technique, it is found that all states are localized in the presence
of the diagonal disorder, in consistence with the main conclusions
of the scaling theory for 2D. In the case of off-diagonal disorder,
however, the states at $E=0$ are delocalized. This delocalization,
inconsistent with the scaling theory and only existing at Dirac
point ($E=0$), is mainly due to the specific chiral symmetry which
is conserved in off-diagonal disorder but destroyed by diagonal
disorder. The obtained conductance at the thermodynamical limit
shows two transparent chirally symmetric quantum channels at $E=0$.
These channels, however, could not produce shape independent
conductivity observed in experiments. It can be obtained only by
introducing the sign randomness of hopping integrals. We investigate
the scaling behavior of the rescaled localization length and the
conductance in the presence of both sign randomness and magnitude
randomness to show the existence of delocalized states at $E=0$.
Especially, the calculated conductivity of a rectangular sample is
shape independent. As the sign randomness of the hopping integrals
corresponds to some additional uncertain phases, the obtained
results reflect a general trend of transition from quantum to
classical transport in dephasing processes. Although the calculated
conductivity is in consistence with the experimental findings, the
mechanism for the phase uncertainty still needs to be identified by
future investigations.

\section*{Acknowledgments}

This work was supported by the State Key Programs for Basic Research
of China (2005CB623605 and 2006CB921803), and by National Foundation
of Natural Science in China Grant Nos. 10474033, 10704040, and
60676056.


\end{document}